\documentclass[epjH,final]{svjour}
\usepackage{graphicx}
\usepackage[utf8]{inputenc}
\usepackage{nicefrac}
\def\degr{\hbox{$^\circ$}}

\def\utw{\smash{\rlap{\lower5pt\hbox{$\sim$}}}}
\def\udtw{\smash{\rlap{\lower6pt\hbox{$\approx$}}}}

\def\farcs{\hbox{$.\!\!^{\prime\prime}$}}

\begin{document}
\title{An international campaign of the 19th century to determine the
solar parallax}
\subtitle{The US Naval expedition to the southern hemisphere 1849 -- 1852}
\author{Andreas Schrimpf\thanks{\email{andreas.schrimpf@physik.uni-marburg.de}}}
\institute{Philipps-Universität Marburg, Fachbereich Physik, Renthof 5,
D-35032 Marburg, Germany}
\def\titlerunning{A. Schrimpf: The expedition to the southern hemisphere
1849–1852}
\date{Received 7 September 2013 / Received in final form 20 December 2013 /
Published online 4 February 2014}
%
%
\abstract{
In 1847 Christian Ludwig Gerling, Marburg (Germany), suggested
the solar parallax to be determined by measuring the position of
Venus close to its inferior conjunction, especially at the stationary points,
from observatories on nearly the same meridian but widely differing in latitude. 
James M.\ Gilliss, 
astronomer at the newly founded U.S.\ Naval Observatory, enthusiastically
adopted this idea and procured a grant for the young astronomical community of
the United States for an expedition to Chile. There they were to observe
several conjunctions of Venus and oppositions of Mars, while the accompanying
measurements were to be taken at the US Naval Observatory in Washington D.C.
and the Harvard College Observatory at Cambridge, USA.
This expedition was supported by A.\ v.\ Humboldt, C.F.\ Gauß, J.F.\ Encke,
S.C.\ Walker, A.D.\ Bache, B.\ Peirce
and others. From 1849 to 1852 not only were astronomical, but also
meteorological and magnetic observations and measurements recorded, mainly
in Santa Lucia close to Santiago, Chile. By comparing these measurements with
those taken simultaneously at other observatories around the world the solar
parallax could be calculated, although incomplete data from the corresponding
northern observatories threatened the project's success. In retrospect this
expedition can be recognized as the foundation of the Chilean astronomy. The
first director of the new National \mbox{Astronomical} Observatory of Chile was
Dr.
C.W.\ Moesta, a Hessian student of Christian Ludwig Gerling's. The exchange of
data between German, American and other astronomers during
this expedition was well mediated by J.G.\ Flügel, consul of the United States
of America and representative of the Smithsonian Institution in Europe, who
altogether played a major role in nurturing the relationship between the growing
scientific community in the U.S.\ and the well established one in Europe at that
time.
} 
\maketitle
\newpage
\section{Introduction}
\label{intro}
In the 17th century determining the size of the earths orbit (the
astronomical unit) and thus, by applying Kepler's law, the size of the solar
system was a matter of realising the dimension of the known universe. After the
surge of Newtonian mechanics in the 18th century the precise astronomical unit
became crucial to understanding the structure of the solar system, e.g.\ the
masses of sun and planets and the perturbations of the Kepler orbits of planets
and small bodies. Finally, in the 19th century the first calculation of a
stellar parallax by Bessel \cite{Bessel1839} showed that the astronomical unit
is a key value in the cosmological distance ladder. It was therefore
unsurprising that Benjamin A.\ Gould, a 19th century pioneer of the
American astronomy stated in his introduction to the analysis of the
above mentioned research campaign, \textit{''The measure of the sun's
distance has been called by the Astronomer Royal of England the noblest problem
in astronomy''} \cite{Gould1856}.

One of the main challenges to solving this problem was the parallax
measurements of the inner planets Mercury and Venus during their transits. After
some first observations in the 17th century, scientists vested all their hope of
accurately determining the dimensions of the earth's orbit into the transits of
Venus, recorded in 1761 and 1769. The first of these transits failed
to improve the solar parallax value, but the results stimulated
the scientists in their efforts for the next transit in 1769.
The values then derived for the solar parallax from different participating
groups ranged from 8\farcs42 to 8\farcs82. It was Franz Encke, who, after a
detailed analysis of all data for both transitions finally deduced a value for
the solar parallax of $8\farcs5776\pm0\farcs0370$ \cite{Encke1824}. One critical
point was that the data delivered by the two most distant stations, at Wardhus
in Norway and at Otaheite on the Friendly Islands, be most accurate. On page
109 Encke says: \textit{''The probable error in the observation of a contact
being seven seconds is so great as to render the determination of the sun's
parallax within 0\farcs01 almost hopeless for the next two centuries. \ldots \
Such precision would require a hundred of observers at Wardhus and the 
same number at Otaheide. \ldots \ All the
observations of 1761 together have but a value equal to three complete
observations at Wardhus and at Otaheite. If the weather would have been good at
the eight northern stations in 1769 and there had been eight good observations
at the Friendly Islands, these sixteen observations would have been as valuable
as the 250 observations of the two transits actually made''} \cite{Encke1824}.
After Littrow discovered that Father Hell, the astronomer at Wardhus in 1769,
had computed an erroneous time for his observations, Encke revised the
computation of the solar parallax value for the transit of 1769 and, combining
the corrected value with that of the transit in 1761, published the value 
$8\farcs57116\pm0\farcs0370$ \cite{Encke1835}, which has been accepted since.

Two other methods had been suggested to find a more precise number for the solar
parallax. Around 1760 Tobias Mayer \cite{Mayer1767}
developed the so called lunar theory and used the motion of the moon to deduce
the solar parallax. One of the perturbation terms depends on the angle between
sun and moon, the coefficient of that term on the solar parallax.
Later, in 1825 Bürg, used a more refined lunar theory to find a value of 
$8\farcs62\pm0\farcs035$ \cite{Buerg1825} and Laplace calculated a slightly
smaller value of 8\farcs61 \cite{Gould1856}. 

One of the earliest ideas of how to find the size of planetary orbits was to
measure the position of Mars with observatories at different latitudes. In the
17th century Cassini and Richer made the first efforts to observe the
declination of Mars. In 1750 Lacaille made more such observations during his
expedition to the Cape of Good Hope and compared them with corresponding
measurements in the northern hemisphere \cite{Gould1856}.
At that time the instruments lacked the necessary precision. 
However, Encke did point out that \textit{''these measurements can be useful
to encourage the hope that by a greater perfection of instruments we may become
more independent of the transits of Venus, which occur so seldom''}
\cite{Encke1824}.

It was Christian Ludwig Gerling who in 1847 came up with a \textit{third method}
(not counting the lunar theory): \textit{the observations of Venus during the 
period of its retrograde motion, especially at its stationary points}
\cite{Gerling1847}.
The paper presented here is based on publicly available information about the
U.S.\ Naval expedition to the southern hemisphere and the scientific estate of
Chr.\ L.\ Gerling containing his correspondence concerning this expedition
\cite{GerlingArchive}.

\section{Christian Ludwig Gerling}
\label{sec_gerling}

\begin{figure}[t]
\begin{center}
\resizebox{0.70\columnwidth}{!}{\includegraphics{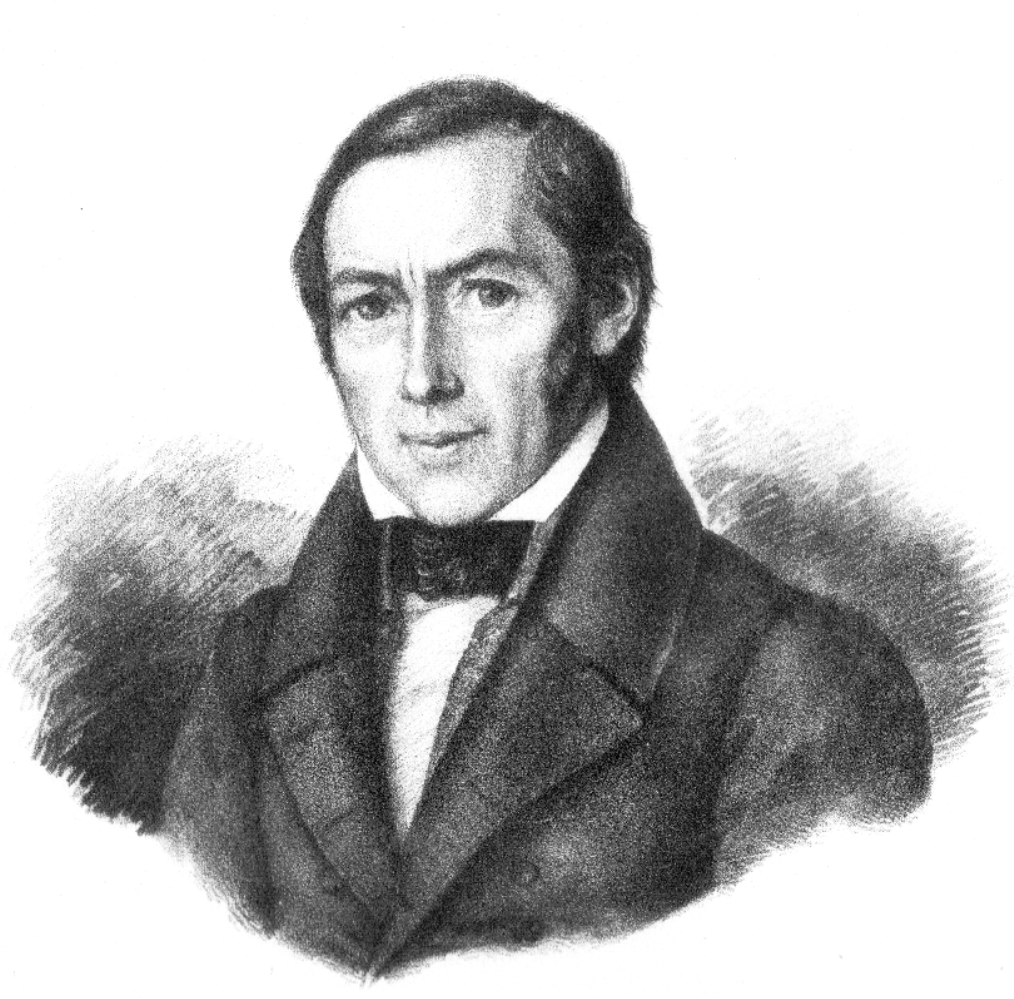}}
\caption{Christian Ludwig Gerling (1788--1864).} 
\label{Fig_Gerling}
\end{center}
\end{figure}

Christian Ludwig Gerling (Fig.\ \ref{Fig_Gerling}) was born in Hamburg, Germany,
in 1788. He was educated together with his longtime friend Johann Franz Encke,
who later became director of the Berlin Observatory. After finishing school
Gerling attended the small University of Helmstedt, but in 1810 he continued his
academic education in the fields of mathematics, astronomy, physics and
chemistry at the University of Göttingen. He started to work at the observatory
of Göttingen under Carl Friedrich Gauß and Karl Ludwig Harding, and, after
some visits in 1811 to the observatories of Gotha (Seeberg), Halle and Leipzig,
he completed his PhD in 1812. 

After he received his PhD Gerling entered a position at a high school in Cassel,
Hesse. At that time he used a small observatory in Cassel for astronomical
observations and occupied himself with calculating the ephemerides of the
asteroid Vesta. He continued to seek a university position and finally in
1817 was appointed full \mbox{professor} of mathematics, physics and astronomy
and director of the ''Mathematisch--Physikalisches Institut'' at the Philipps--Universität of
Marburg.
In spite of several offers he remained at the university in Marburg till his
death in 1864 \cite{Madelung1996}.

Gerling's scientific work was affected by two major topics: in his early
period in Marburg from 1817 to 1838 he was well occupied with organizing
the triangulation of Kurhessen. During that time Gerling did not have an
observatory, so his astronomical work was reduced mainly to corrections
on a section of Encke's ``Berliner Akademische Sternkarten''. 

In 1838 the institute moved to a new home in Marburg at the
''Rent\-hof''. After the building was reconstructed in 1841, he
could finally put into operation his new but small observatory, built on top of
a tower of Marburg's old city wall \cite{Schrimpf2010}.
Gerling pursued the scientific topics of astronomy of that time, making
meridional observations and differential extra-meridional measurements of stars,
planets and asteroids, observations of lunar occultations, etc.\ mainly to
improve the precision of star catalogs and orbit parameters of solar system
bodies.

Carl Friedrich Gauß and Christian Ludwig Gerling started their relationship as a
teacher and his student but during the following years they became each others
counselor and finally close friends. Their correspondence not only contains
details of scientific discussions but also reflects their close relationship
$[$Schäfer 1927;Gerardy 1964$]$
%
Gerling died in 1864 at the age of 76. His estate has been
stored in the library of the Philipps--Universität Marburg \cite{GerlingArchive}.
About 1000 letters of his correspondence with Gauß, Encke, Nicolai, Gilliss,
Flügel and many more give an insight into his personal and professional life
and can be a valuable source for historians.

\section{The idea of the project}
\label{sec_idea_venus_measurements}
In 1841 Christian Ludwig Gerling was able to begin operating his new
observatory.
The discovery of new asteroids, beginning with (5) Astrea in 1845 and the
expansion of the solar system after the discovery of the planet Neptune in 1846
must have encouraged many astronomers to increase their research
into the solar system.

A clue to the ideas that occupied his time can be found in the letters
that Gerling and Gauß exchanged \cite{Schaefer1927}. They encouraged each
other to observe \textit{''Leverriers planet''} (\cite{Schaefer1927}, pp.\ (740
and 742) as they called Neptune, but there is no hint at any special focus of
Gerling's on the solar parallax. However, in 1847 Gerling published his article
\textit{''About the Use of the Venus Stationary Points for Determination of the
Solar Parallax''} \cite{Gerling1847}  (title translated by the author).

\begin{figure}[ht]
\begin{center}
\resizebox{0.79\columnwidth}{!}{\includegraphics{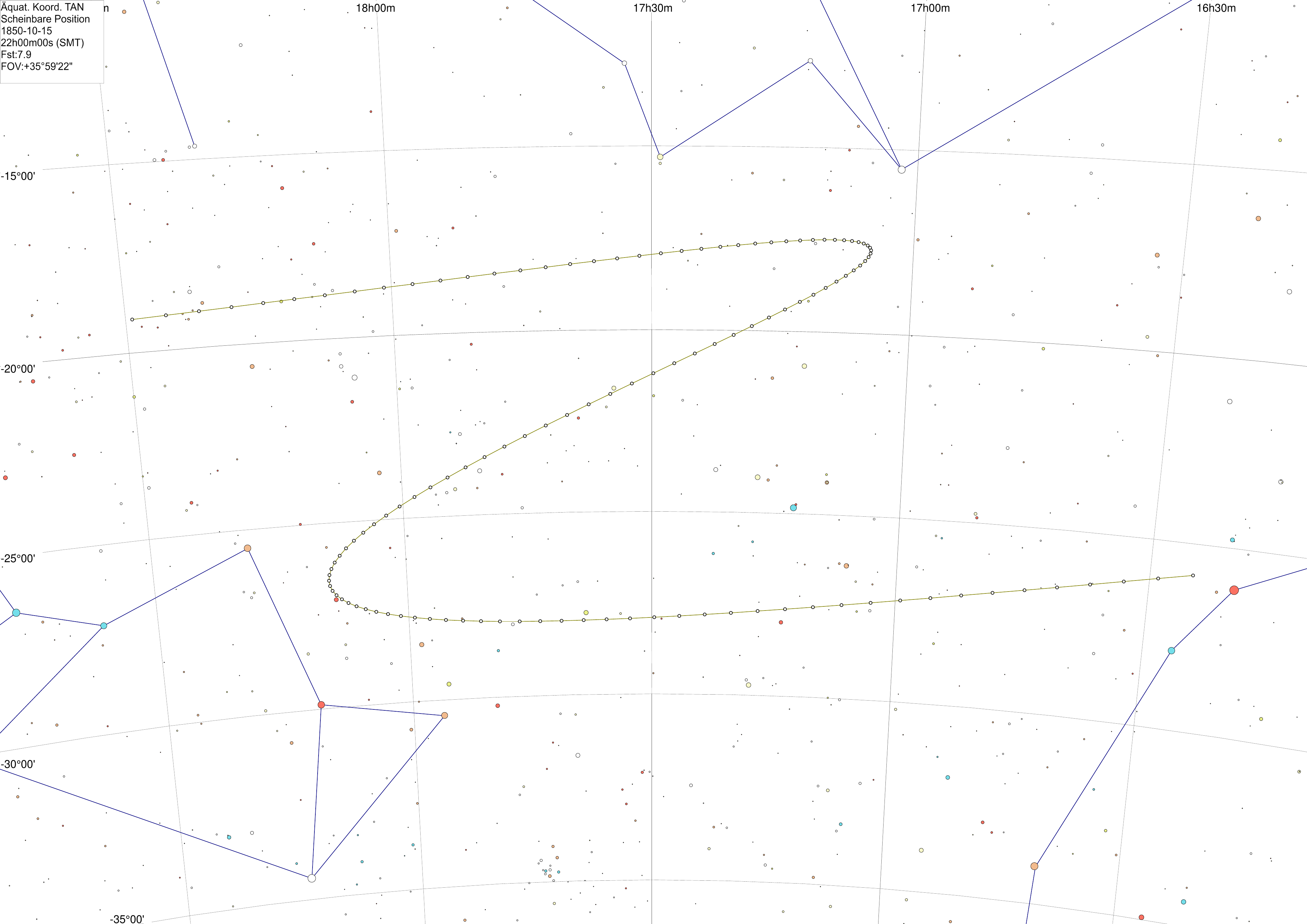}}
\caption{The loop of the planet Venus in 1850/1851 according to JPL ephemerides
DE423. 
Gerling suggested measuring the position of the planet through differential
meridional and extra-meridional observations, especially at the
stationary points of the loop, where the apparent motion of the planet is almost
zero. He calculated a rough observation plan for the Venus loops of 1847 and
1849.
However the final observations were made during the loop of 1850/1851.}
\label{Fig_Venus_Loop_1850}
\end{center}
\end{figure}

There are two main points in this paper: First, Gerling suggests performing
parallax measurements of Venus instead of Mars, because during the inferior
conjunction, Venus is much closer to Earth than Mars is during its opposition,
which thus increases the observed planet's parallax and reduces the error of
the solar parallax derived. There is, however, a disadvantage in observing Venus:
during the conjunction it is visible only at day time and only few
bright stars can serve as comparisons for differential measurements. Under
these conditions observatories required meridional instruments of highest
quality to obtain results, which could improve the solar parallax value.
Second, Gerling proposes observing Venus at the stationary points of its loop
(Fig.\ \ref{Fig_Venus_Loop_1850}), for two reasons. One of these is that at
these points the distance between the sun and Venus is about 29\degr and, using
equatorial instruments, Venus can easily be observed early or late at night with
many comparison stars. The other reason is that at the
stationary points the planet is moving so slowly that day-time meridional
observations at different latitudes can easily be reduced by interpolation to
the same meridian, if the longitudes of the observatories involved are fairly
similar. Higher precision can be achieved by increasing the number of
observations.

On the July 15th, 1847 Gauß acknowledged the interesting article
(\cite{Schaefer1927}, letter no.\ 364), but he was unconvinced that comparing
meridional with equatorial measurements of lower quality would lead to an
improvement of the solar parallax. Two days later Gerling sent his thanks to
Gauß and, noting his concerns, he added that the same comparison
stars used for differential measurements of Venus at its stationary points should have to
be observed with meridional instruments when they are accessible earlier in the
year (\cite{Schaefer1927}, letter no.\ 365). Gerling also published a statement
with these additional explanations in the \textit{Astronomi\-sche Nachrichten}
\cite{Gerling1847b}.

The success of the project required the cooperation of suitable
observatories in different hemispheres of the world. In his publication 
\cite{Gerling1847} Gerling considered observatories in Königsberg and the
Cape of God Hope, in Washington and the Antilles, and of course in Greenwich. 
Gerling began collecting information about the situation of astronomy in
different parts of the world. In his pursuit he contacted Dr.\ J.\ Flügel, the
consul of the United States in Leipzig, Germany, to learn about the situation
of astronomy in the United States (\cite{GerlingArchive}, Ms.\ 319/217, Flügel
to Gerling, February 8th 1847). He also sought to directly contact
someone in South America, and although this attempt was in vain
(\cite{GerlingArchive}, Ms.\ 319/219, Flügel to Gilliss, May 15th 1847),
Gerling was perhaps already thinking about a new observatory in that part of
the world.

\section{James Melville Gilliss}
\label{sec_gilliss}
Before moving on to the realization of Gerling's idea, one must take a closer
look at James Melville Gilliss, the key person in organizing and leading the
expedition to the southern hemisphere. A good source of information about his
life can be found in his memoir, read before the National Academy on January
26th, 1866 by Benjamin A.\ Gould \cite{Gould1866}.

Gilliss was born in 1811 and entered the U.S.\ Navy at the age of 15. Being
interested in scientific work, he applied for leave of absence in 1833 to pursue
an education at the University of Virginia. In 1835 he continued his studies in
Paris before returning to his professional duties as a lieutenant of the Navy in
Washington. He was given responsibility for the care and distribution of charts
and instruments required by national vessels and for the rating of chronometers
by determining the time with transit instruments. In 1838 he joined the U.S.\
Exploring Expedition as the observer responsible for moon culminations,
occultations and eclipses to determine differences in longitudes. Gilliss
observed more than 10.000 transits, embracing the moon, the planets and about
1100 stars, and finally compiled the first report of astronomical observations
conducted by a US scientist. B.A.\ Gould stated:
\textit{''Gilliss must be considered the first representative of practical
astronomy in America. It was Gilliss who first in all the land conducted a
working observatory, it was he who first published a volume of observations,
first prepared a catalogue of stars''} \cite{Gould1866}. After his return from
the expedition the Secretary of the Navy assigned him the duty of preparing
building plans and collecting the instruments for a new
observatory.
In September 1844 Gilliss reported the work complete and the observatory ready
for use. However, it was Lieut.\ M.\ F.\ Maury, a young officer without
scientific education or experience, who became the first director of the new 
U.\ S.\ Naval Observatory in Washington. Maury had become known for his
contributions to charting winds and ocean currents, but never as an astronomer.
Gilliss was assigned to duty upon the Coast Survey under Professor Bache.
 
It is very likely that Dr.\ J.\ Flügel made Gilliss and Gerling known to each
other. Gilliss probably came into contact with Flügel while he was
planning the instruments for the U.S.\ Naval Observatory \cite{Gould1866}.
In turn, Flügel sent Gerling a volume of Gilliss' observatory report. After
Gerling had a look at this report, he published a very positive note in the
\textit{Heidelberger Jahrbuch} of 1847 and in August 1847 Gilliss was elected a
member of the \textit{Naturwissenschaftliche Gesellschaft}
(\cite{GerlingArchive}, Ms.\ 319/220, Flügel to Gerling, August 30th 1847).

In June 1847 Gilliss received a letter from  Gerling introducing the idea of
the Venus observations for determining the solar parallax. Gilliss was
overwhelmed by this proposal, which should almost completely occupy his life
henceforward. From 1849 to 1852 he operated the expedition to the southern
hemisphere and after returning began with reducing the data. In 1855
the Congress of the U.S.\ enacted a law, that Navy officers
without recent active duty on board of a Navy vessel, should be put on a
''reserved list'' and retired. Although Gilliss found his name on this
list, the Secretary of the Navy offered him the same salary to continue
his work and prepare the remaining volumes of his report. Gilliss was later
returned to duty, in 1861 he received a commission as Commander and a year
later as Captain (\cite{GerlingArchive}, Ms.\ 318/295, Gilliss to Gerling,
November 23rd 1855, and \cite{Gould1866}).
In 1858 and 1860 he went on expeditions to observe total eclipses. 
At the beginning of the American Civil War Maury left the U.S.\ Navel
Observatory and finally in 1861 Gilliss was set in charge of this
institutions, \textit{''a post which the scientific world had expected for him
sixteen years before''} \cite{Gould1866}. He revived the idea of determining
the solar parallax by simultaneous observations in Chile and in the United
States --- and finally succeeded! He died early in 1865 at the age of 54, just
as the results of these observations were published.

During the years they were communicating, Gilliss and Gerling became friends,
sharing personal matters of their professional lives and each others' family
affairs. However, neither did Gilliss return to Europe nor Gerling travel to the
United States; they never met personally.

\section{The expedition to the southern hemisphere}
\label{sec_expedition}
A letter of Gerling's to Gilliss dated 17th April 1847 marked the beginning of
the expedition to the southern hemisphere \cite{Gilliss1856}\footnote{Gilliss
included the organization and operation of the expedition, the
observations taken in Chile and the data analysis in volume III of his report
to the House of Representatives. There were however many other very important
observation and measurements, which Gilliss reported in volumes I, II and VI.
The volumes IV and V never were published. The data planed to be included
in these volumes where published later by the U.S.\ Naval Observatory in its
Astronomical and Meteorological Observations. (Washington observations) for
1868, app.\ I, 1871, and Astronomical, Magnetic and Meteorological observations
(Washington observations) for 1890, app.\ I, 1895.}.
In this letter Gerling explained the idea of the specific Venus
observations including some first information about the Venus loops of 1847 and
1849, hoping to attract Gilliss' attention and through him that
of the American astronomers. Gerling suggested the co-operation
of many observatories in the northern part of the world with a few meridional
instruments in the southern hemisphere, which \textit{''perhaps is in your
{\normalfont (Gilliss)} power to facilitate''} (\cite{Gilliss1856}, p.\
iv).

It is remarkable that at that time the young and developing American
\mbox{scientific} community was becoming involved in astronomical research but
had not yet \mbox{published} a noticeable contribution to science
\cite{Rothenberg1999}. Gerling's letter to Gilliss there\-fore came at an
opportune time: new observatories had opened in the United States and the
American astronomical community was waiting for chance to participate in the
progress of science.
Gilliss' very first reaction was to pass Gerling's letter on to Prof.\
Sears Cook Walker (at that time at the U.S.\ Naval Observatory) and soon both
were forwarding it to all American astronomers and observatories and to the
\textit{National Intelligencer}, one of the leading newspapers in Washington at
that time. Noteworthy are the several articles this newspaper published during
the preparations and actual course of the expedition.

Until November 1847 Gilliss studied the idea. Counting on the new U.S.
Naval Observatory in Washington, the Harvard College Observatory, founded in
1839, and the observatories in Philadelphia and at West Point as possible
northern stations, and on a new observatory on the island of Chilo\'e in the
South of Chile as a southern station, Gilliss, the visionary, foresaw the
relevance: \textit{''I am the more keenly sensible of the noble opportunity now
within our grasp to present the world, from our own continent as a base, the
dimensions of our common system''} (\cite{Gilliss1856}, p. v).

\subsection{Planning the expedition}
\label{subsec_expedition_planing}
In the fall of 1847 Gilliss and Gerling were busy collecting the opinions of
well known astronomers. C.F.\ Gauß, though not really convinced that Gerling's
method could result in a more precise solar parallax, did suggest testing the
idea 
([Schäfer 1927, letter no.\ 368]) 
whereas Prof.\ Boguslawski, director of
the Breslau observatory welcomed and published it in his circular
\textit{Uranus}. F.\ Encke also sent an encouraging note.

Gilliss soon was able to convince all of the leading American astronomers! After
receiving Gerling's forwarded letter Prof.\ A.D.\ Bache replied:
\textit{''I do not see, however, that the two reasons which strongly favor
Dr.\ Gerling's method are met by any opposing arguments. The large number of
observations upon which results may be found, and the independence of the new
method with that formerly used, are, indeed, striking factors in this
method''} \cite{Gilliss1856}.

B.\ Peirce, professor of mathematics at Harvard University, responded:
\textit{''A more accurate measurement of the sun's parallax in the the method
proposed by yourself ({\normalfont Gilliss}) and Dr.\ Gerling cannot be
regarded as inferior in importance to any problem in practical astronomy. \ldots
Most cordially, therefore, as well as deliberately, I send you my humble
testimony in favor of the enterprise you have so much at heart and are so
certain to accomplish, if it is, as it appears to be, within the bounds of
reasonable possibility''} \cite{Gilliss1856}. And then Peirce continued:
\textit{''I shall bring the subject before the Academy ({\normalfont i.e.\ the
Academy of Arts and Science}) early in the month of January ({\normalfont
1848}), and shall be greatly disappointed if a resolution is not immediately adopted in
approbation of your project''} \cite{Gilliss1856}.

In December 1848 Gerling relayed the opinions of Gauß, Encke and Boguslawski
and discussed the location of the new southern observatory. Unlike Gilliss he
preferred an observatory on the mainland of Chile and not at one of the southern
islands. He recommended a location near to Valparaiso, mainly because of
the more stable weather conditions in that area, whereas the closer proximity
to the northern station at Washington D.C. could be compensated by a much higher
number of expected observations. Gilliss then also agreed to  Valparaiso.

In his reply to Prof.\ Peirce Gilliss estimated the expected error of the
solar parallax reduced from the proposed measurements to be smaller than
0\farcs1. He proposed an expedition to a location near Valparaiso to
observe Mars - in the same way as former observations of Mars have been done
- and to make a new attempt at observing Venus during her retrograde
motion.

This proposal was put forward to the two leading scientific organizations of the
country, the Academy of Arts and Science and the American Philosophical
Society. The Secretary of the Navy referred the matter to the action of
Congress, and F.P.\ Stanton, the chairman of the Naval Committee cordially
approved the plan, limiting the expenses for the expedition to \textdollar{}
5000. With this budget it passed both Houses of Congress and was finally
approved by the President of the United States on August 3rd 1848.

Soon after the news reached Gilliss, he started with detailed preparations
for the expedition. He proposed observations of
\begin{itemize}
  \item the Mars oppositions of 1849 and 1852 using simultaneous
  extra--meridional measurements taken at pre-arranged times in
  participating observatories to determine the parallax in right ascension and
  using meridional measurements for the parallax in declination;
  \item Venus at about 
  the times of the inferior conjunctions with the sun in 1850 and 1852 
  and, more particularly, near its stationary terms with
  meridional and extra-meridional differential measurements. That is, daily
  micrometrical comparisons with preselected stars and simultaneous pre-arranged
  comparisons during daylight with bright stars should be made and compared
  between observatories lying near the same parallel for parallax in right
  ascension and observatories at nearly the same meridian but in opposite
  hemispheres;
  \item the moon, smaller
  planets, and lunar occultations;
  \item stars in zones between the south pole and 30\degr of
  south declination;
  \item the terrestrial refraction;
  \item comets;
  \item magnetic and meteorological occurrences;
  \item seismometer evidence of earthquakes.
\end{itemize}

\noindent
The instrumentation required comprised a meridian-circle, an
achromatic equatorial telescope, a sidereal clock and several chronometers, as
well as further instruments for meteorological and magnetic measurements.

\begin{figure}[t]
\begin{center}
\resizebox{0.59\columnwidth}{!}{\includegraphics{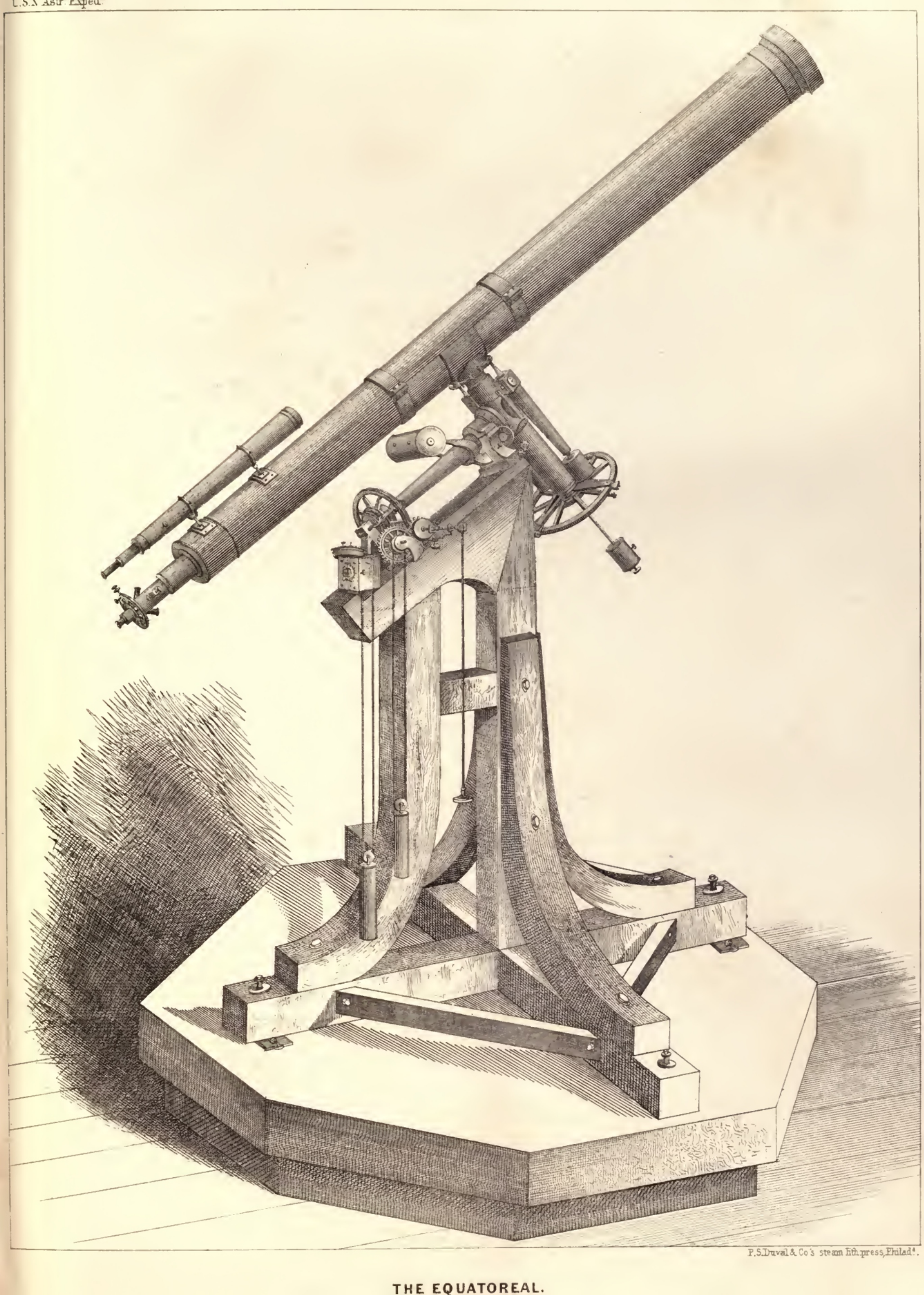}}
\caption{The equatorial of the expedition was the first American telescope of
considerable size! The lens had a diameter of 173 mm and a focal length of 2.63
m and was prepared by Henry Fitz, New York; the mechanics were constructed
by William J.\ Young, Philadelphia \cite{Gilliss1856}.}
\label{Equatorial_Southern_Expedition}
\end{center}
\end{figure}

The meridian-circle with 91 cm (36 inches) in diameter and reading at least to
1'' by four microscopes was ordered from Pistor \& Martins, Berlin.
Because the circle was quite expensive, Gilliss sought a manufacturer in the
States to build the equatorial. He finally succeeded: the mechanics were made by
William J.\ Young, Philadelphia, and the objective was prepared by Henry Fitz, jr, New
York. The diameter of the object glass was about 173 mm (6.4 french inches) with
a focal length of 2.63 m (103.7 inches). This actually is the first American
telescope of considerable size (Fig.\ \ref{Equatorial_Southern_Expedition}), and
\textit{''it marked an era of progress of mechanics and optics in the U.S.\ of
the 19th century''} \cite{Gould1866}!

The U.S.\ Navy was to supply the remaining instruments, but everything
Maury offered to Gilliss was broken, oxidized and thus totally
unusable. Gilliss seemed to be running out of money when he received generous
support from Prof.\ Henry, the Secretary of the Smithsonian Institution. Henry
lent the expedition a seismometer and a complete meteorological outfit.

Gilliss also had to organize the observatory's construction. Small assembly
parts suitable for transportation were set up for inspection in Washington and
then packed for shipment to Chile.

Gerling and Gilliss were in close contact during that time and Gerling regularly
sent a progress report of the preparations to C.F.\ Gauß. Gauß
expressed his compliments toward Gilliss:
\textit{''The American expedition to Chile will surely bring 
some treasurable results. Mr.\ Gilliss has planned to study so many different
topics, that I have nothing to add''} 
([Schäfer 1927, letter no.\ 368],
translated by the author).

\subsection{The expedition to Santa Lucia, Chile}
\label{subsec_expedition_st_lucia}
The departure was scheduled for June 1st 1849, but there 
was still one important matter to take care of: the star charts, including
the comparison stars for the proposed Venus and Mars observations, had to be
prepared. Two former midshipmen, Archibald MacRae and Henry C. Hunter, were
assigned to assist Lieutenant Gilliss. They made selections of stars from
available recent catalogues:
\textit{Mittlere Oerter von 12.000 Fixster\-nen} of Rümker, Hamburg;
\textit{Histoire C\'eleste} of Lalande, Paris; the as yet unpublished
observations of the Washington Catalogue, as well as Bessel's zones. They also
calculated the ephemerides of the two planets and reduced the catalogue entries
to the 1st of January of the year the ephemeris was calculated. These locations
and those of the planets were plotted to charts. A \textit{Circular to the
Friends of Science} was prepared, which explained in short the purpose of
the expedition, the planned observations in Chile and the necessary accompanying
observations at the other observatories. The astronomers in charge were asked to
co-operate and to conduct all the requested observations.

The circular, including copies of the charts, was sent to every known
observatory and it was published in the \textit{Monthly Notices of the Royal
Astronomical Society} \cite{MNRAS1849}.
As Gilliss wrote in his report: \textit{''In order to insure the
impartial trial of Dr. Gerling's  method, under date of August 11, Lieut.\ Maury
was instructed by the honorable Secretary, that - 'As the success of the
Astronomical Expedition to Chile, under the direction of Lieut.\ Gilliss, will
greatly depend on the care with which the corresponding
observations are made in the northern hemisphere, you will designate an
assistant whose special duty it shall be to make the observations at the times
and in the manner specified in the 'Circular to the Friends of Science,' which
you prepared under the direction of this department' ''} (\cite{Gilliss1856},
p.\ xxx).

The equipment was shipped in June on board the \textit{Louis Philippe} to
Chile and on the 16th of August, James Melville Gilliss left New York aboard
the steamer \textit{Empire City} towards Valparaiso, Chile. Gilliss reached
Valparaiso on October 25th, 1849 and travelled through the night to arrive in
Santiago the next day.
There, presenting his letters, he was welcomed by the government:
\textit{''Indeed, the good will and liberality of the President and his Cabinet
then, and throughout our stay in the country, were uniformly manifested''}
(\cite{Gilliss1856}, p.\ xxxi). Fourteen days later Gilliss decided to set
up the observatory at Santa Lucia, a small porphyritic knoll in the eastern
quarter of the city.

B.\ Gould praised the support of the Chilean government: \textit{''The Chilean
\mbox{government} received the expedition with a cordial hospitality, placing at
his disposal any unoccupied public ground, admitting free of duty all effects of the
officers as well as the equipment of the expedition, and from first to last
facilitating the enterprise by every means in their power''} \cite{Gould1866}.

On December 5th, 1849 the observatory building was ready for the
instruments, the equatorial was mounted on the 6th of December and four days
later the series of Mars observation commenced. By the following February the
circle was ready for use and after completing the first series of Mars
observations the zone observations were instituted. \textit{''Each night's
work comprised observations of level, nadir point, collimation by reflection of wires
from mercury and standard stars before and after a zone extending through three
to four hours in right ascension, so that we were occupied from five to six, and
sometimes more hours''} (\cite{Gilliss1856}, p.\ xxxiii). The expedition
was at work!

\begin{figure}[t]
\begin{center}
\resizebox{0.76\columnwidth}{!}{\includegraphics{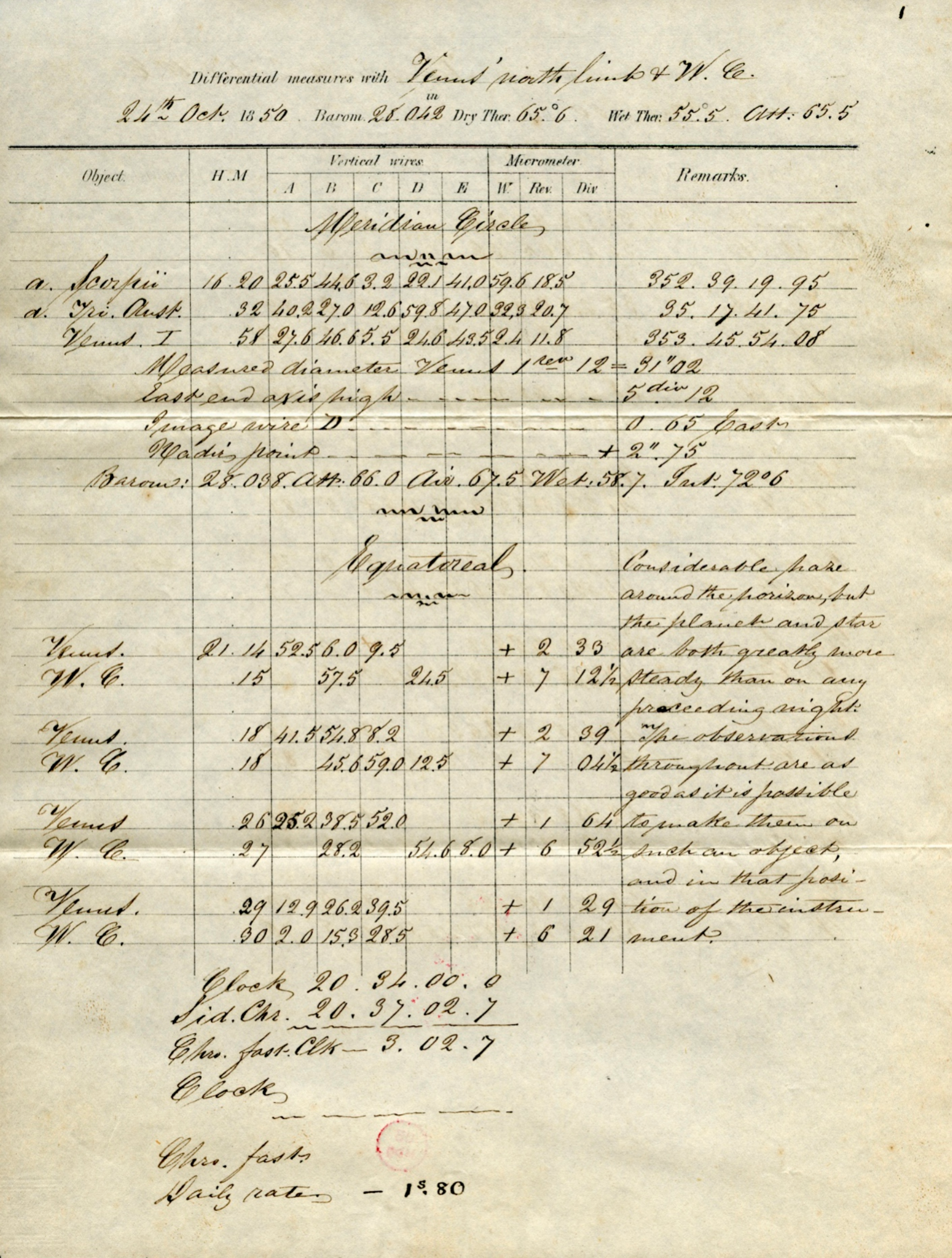}}
\caption{Data sheet of the expedition. Gilliss sent his observations 
of October 24th, 1850 to Gerling (\cite{GerlingArchive}, Ms.\ 319/282). In them
he noted the meridional observations of $\alpha$ Scorpii (Antares),  $\alpha$
Triangulum Australis (Atria) and Venus, 35 differential equatorial observations
of Venus and a of star from the Washington Catalog (W.C.), which could be
identified in the modern UCAC catalog as UCAC4-314-105903, a star with m$_{\mbox{v}}$ =
9.37 mag. Here only the first of five pages is shown. These measurements are
listed in Gould's data reduction (\cite{Gould1856}, pp.\ clxxvii and
clxxviii).}
\label{Venus_Observations_1850}
\end{center}
\end{figure}

\begin{sloppypar}
During the expedition Gilliss maintained close contact to Gerling. On 
October 25th, 1850 he sent him a short report about the progress in Santa
Lucia (\cite{GerlingArchive}, Ms.\ 319/282)
mentioning that Henry Hunter had been injured right at the beginning of
their stay in Chile and was unable to participate in the \mbox{observations}.
The same month a new assistant, Seth Ledyard Phelps, arrived at Santa Lucia.
Gilliss thanked Gerling for a re-computation of some of the comparison
stars for the Mars observations in January 1850, sent him the lunar occultation
observation of a smaller group of stars close to $\chi^2$ Ori and asked him to
have students reduce these data. Knowing that Gerling would be very interested,
he also sent him the Venus observation data of the 24th of October, for an
insight into the conditions and the quality of the data (Fig.\
\ref{Venus_Observations_1850}.)
\end{sloppypar}

The 24th of October was the fifth day of the first Venus series. During
the daytime the astronomers observed the planet Venus and two bright stars --
$\alpha$ Scorpii (Antares) and $\alpha$ Triangulum Australis (Atria) -- 
passing in the meridian circle. At night between 9:20 pm and 11:00 pm they
measured 35 differential equatorial positions of Venus and the comparison star
116 (\cite{Gould1856}, pp.\ clxxvii and clxxviii). In his letter, Gilliss
explained the conditions at night which allowed him to make between 30 and 40
observations before Venus and the comparison star were set.

On the clear nights, while he was using the equatorial, the assistants spent a
lot of time with the meridian circle, adding an average of more than a thousand
stars to the catalogue each month. They checked their personal estimates of
magnitudes by comparing their observations of small and well-known stars with
the British \mbox{Association} catalogues. They also observed the famous
variable star $\eta$ Argus, now $\eta$ Carina (\cite{Gilliss1853}, and see next
subsection), which, through observations by Halley, Lacaille, John
Herschel and others, had been known to display huge an irregular changes
in brightness. Their observations of $\eta$ Carina had been recorded between
February 1850 and March 1852, with $\eta$ Carina reaching the brilliance of
Canopus and topping that of the two stars of $\alpha$ Centauri.

The time in Chile was rapidly drawing to a close. Gilliss, MacRae and Phelps
had made in total 217 series of observations of Mars and Venus, extending over
nearly three years. They had recorded about 7000 meridional observations of 2000
stars, mainly standard stars which were used for determining instrumental
errors. Their monitored zones comprised approx.\ 33.000 observations of about
23.000 stars within 24 $\nicefrac{1}{5}$\degr\ of the south pole.
\textit{''Such are the astronomical results of this most honorable and useful
expedition''}, Gould stated in his memoir of Gilliss 
\cite{Gould1866}.

On January 5th, 1852 they were able to observe a very rare
double occultation of $\eta$ Geminorum at the observatories in Valparaiso and
Santa Lucia \cite{Gilliss1852} and to synchronize their clocks via telegraphic
signals. They could thus determine the difference in longitude between the two
locations to be 3 m 56.51 s $\pm$ 0.021 s \cite{Gilliss1856} in response to
Alexander von Humboldt, who had asked for new data about the longitude of
Valparaiso (see next subsection.)

Lieutenant MacRae was sent home via the Uspallata Pass of the Andes, where he
made some more meteorological and magnetic measurements. He returned to the
States in April 1853. Phelps and Gilliss left Valparaiso by steamer on the
1st of October and reached New York in November 1852.

\subsection{Acknowledgements of the expedition}
\label{subsec_expedition_acknowledgements}
Dr. Flügel, the consul at Leipzig, acknowledged receipt of the circular in a
letter to Gerling, dated September 18th 1849. He distributed the
circular among many of his contacts and reported back the responses. One of them
came from John Parish, Baron of Senftenberg, in the Pardubice Region of the Czech Republic. 
John Parish had built a private observatory at his castle and employed the
Danish astronomer Theodor Brorsen. He offered to assist in the accompanying
observations but regretted that his smaller instrument might lack the precision
needed to observe the smaller stars
(\cite{GerlingArchive}, Ms.\ 319/232, Flügel to Gerling, December 31st
1849). Unfortunately, no observations of the Senftenberg observatory
were mentioned in Gilliss report.
A short note also arrived from Prof.\ Kupffer, St. Petersburg
(\cite{GerlingArchive}, Ms.\ 319/233, Flügel to Gerling, February 4th 1850),
saying that he would contact Gerling directly, although so far no letter of
Prof.\ Kupffer could be found in the Gerling--Archive.

\begin{figure}[t]
\begin{center}
\resizebox{0.75\columnwidth}{!}{\includegraphics{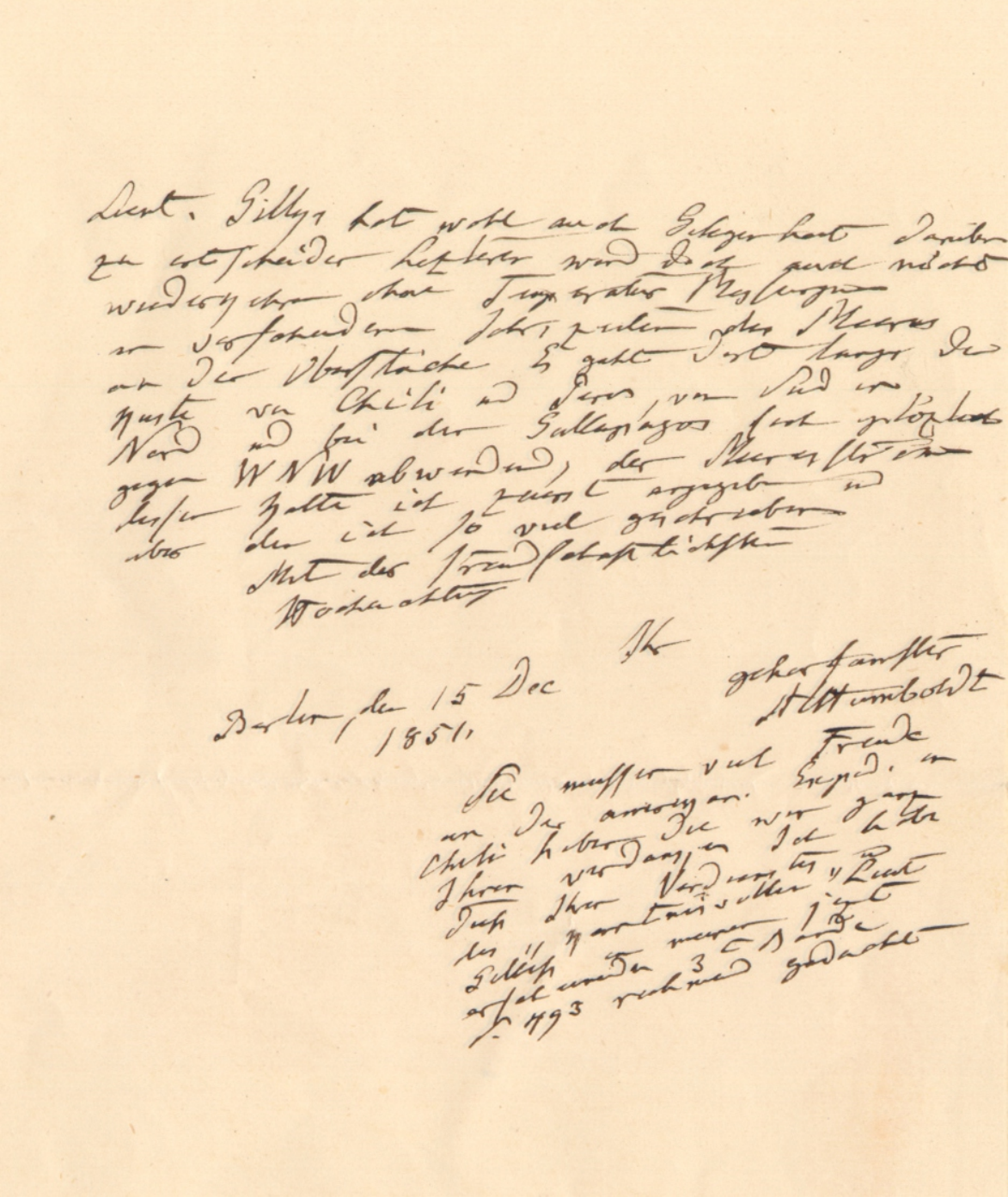}}
\caption{Last page of a letter from  A.\ v.\ Humboldt to Gerling dated
December 15th 1851 (\cite{GerlingArchive}, Ms.\ 319/362). He asked Gerling
to send a note to Gilliss in Chile and have him perform some additional
observations. In the postscript Humboldt mentioned the acknowledgement of the
southern expedition in his book \textit{Kosmos}, vol.\ 3 \cite{Humboldt1850}. }
\label{Fig_5_Letter_AvH}
\end{center}
\end{figure}

Another very interesting response came from Alexander von Humboldt, dated
December 22nd 1849 (Fig.\ \ref{Fig_5_Letter_AvH}):
\textit{''I hasten, my respected Doctor, to express to you my best thanks
\ldots for the notices of Gilliss referring to the ascertainment of the parallax in
Chili, and the astronomical longitude of Washington. \ldots \ For an
(antediluvian?) who, like myself, is attached with his whole soul to the new
world \ldots \ it is renovating and pleasing to follow the quick and proud
expansion of scientific sentiments in the United States, and to have to
acknowledge how much the government participates in an expedition to Chili of
three years, because of a Professor in Marburg wishing for it, and nobody
listens to him in Europe!''} (\cite{GerlingArchive}, Ms.\ 319/232, Flügel sent a
copy of Humboldts letter to Gerling, and \cite{Schwarz2004}, letter no.\ 156).

Gilliss kept up his correspondence and on February 25th, 1850 he sent a
letter to Flügel about their observation of the variable star $\eta$ Argus (now
$\eta$ Carinae), which had become very famous after John Herschel's observation
of the dramatic change in magnitude. Flügel passed this letter on to Humboldt,
which spurred his interest in this expedition (\cite{Schwarz2004}, letter
no.\ 160). In December 1851 Humboldt contacted Gerling concerning the
expedition and expressed interest in two observations. First, he asked
that Gilliss take a good measurement of the longitude of Valparaiso, to
compare it with the reduction from his observation of a Mercury transit. Second,
he was interested in a red glow, that he, Humboldt, had observed on high snow-covered
mountains during sunrise and sunset in the Alps and the Himalayas. He asked
if Gilliss had information of such a red glow of the Andes. He ended his
letter with the very cordial comment: \textit{''You must be very pleased about
the American expedition to Chili, which we owe solely to you. I have praised
yours and the very knowledgeable Gilliss' achievement in my recently published
book 'Kosmos', volume no.\ 3, page 493''} (\cite{GerlingArchive}, Ms.\ 319/362, and
\cite{Schwarz2004}, letter no.\ 177, translated by the author).

\section{The results of the expedition}
\label{sec_results}
After returning home Gilliss was occupied with reducing the observations and
preparing the reports of his expedition for the Navy Department.

In 1855 the first volume, a narrative report about Chile, was published.
However, Gilliss, fearing he would not be able to focus his full attention
on determining the solar parallax, arranged for the total body
of the data to be given to Benjamin A.\ Gould, at that time director of the
Dudley Observatory, for further analysis.

Unfortunately, very disappointing news came soon from the northern
observatories: the Navel Observatory at Washington reported eleven observations
of Mars, three of which were partially unsatisfactory, and eight of Venus,
two described as poor. The Harvard Observatory at Cambridge reported five
observations of Mars, four of them of one limb only, and the Royal Observatory
at Greenwich could \mbox{report} four observations of Mars, three of which were
considered poor! Gould could not contain his disappointment:
\textit{''These are the only observations at hand for combining with the
magnificent series of Lieutenant Gilliss, according to the method suggested by
Professor Gerling, and contemplated by the expedition. It is impossible to
refrain from the expression of deep regret that, from all the observations of
the well equipped and richly endowed observatories of the northern hemisphere,
so few materials could be found''} \cite{Gould1856}. To some extent the rivalry
between Gilliss and Maury --- in spite of the order Maury received --- may 
account for the failure of the northern observations. Why the observers
at Cambridge did not follow the original and euphorically accepted
schedule of measurements remains open.

As the corresponding observations were totally inadequate, Gould had to seek
other observations made in each hemisphere during the Mars oppositions and
Venus conjunctions between 1849 and 1852. Besides the meridional
observations and the micrometric comparisons from Santiago, Washington,
Cambridge and Greenwich, Gould received data from the Cape of Good Hope
(micrometric comparisons and meridional observations of Mars), Athens
(meridional observations of Mars), Cracow (meridional observations of
Venus), Kremsmünster (meridional observations of Mars), and Altona (meridional
observations of Venus).

For each of the comparison measurements Gould calculated the absolute position
and compared these with the meridional observations. These results were combined
in series of approximate solutions by means of least squares. He found
that the results from the first Mars opposition by far surpasses the three other
series in precision; he therefore used this series for further analysis.
First, Gould noted the results for 
the angular semi-diameter in a mean distance of one astronomical unit of the two
planets, determined solely from observations of Santiago and Washington
\[
\begin{array}{lcl}
\mathrm{semi-diameter\ of\ Mars} &	= & 6\farcs83\\
\mathrm{semi-diamater\ of\ Venus}&  = &	8\farcs35.
\end{array}
\]
and then gave the results for solar parallax
\[ \mathrm{mean\ solar\ parallax} = 8\farcs4950. \]

\vspace{2em}
\noindent
Comparing this with the slightly higher value obtained by Encke he
suggested adopting 8\farcs5000 as a more accurate value for the solar parallax.
He published these \mbox{results} in \textit{The Astronomical Journal}
\cite{Gould1858}, Gilliss in the \textit{Monthly Notices of the Royal
Astronomical Society} \cite{Gilliss1859} and Gerling in \textit{Astronomische
Nachrichten} \cite{Gerling1858}.

\section{Some impacts of the expedition}
\label{sec_impacts_expedition}
\subsection{The beginning of the Chilean astronomy}
\label{subsec_beginning_chilean_astronomy}
\begin{sloppypar}
The Chilean government was pleased with the astronomical activities in 
\mbox{Santiago.}
Early in 1850 they asked Gilliss to instruct three young Chileans in
practical astronomy, with the intention of establishing a permanent national
observatory \cite{Gerling1851}.
\end{sloppypar}

In October 1850 Dr.\ Carl W.\ Moesta, a former student of Gerling's, came
to Chile and introduced himself to Gilliss, showing  Gerling's certificate of
qualification. He received an appointment as Professor in the College at
Coquimbo (\cite{GerlingArchive}, Ms.\ 319/282, Gilliss to Gerling, October
25th 1850).
A year later Gilliss helped Moesta attain a post as assistant to the chief of
the Topographical Survey in Chile, a piece of news that Gerling was happy to
send to Gauß (\cite{Schaefer1927}, letter no.\ 376).

As the expedition was drawing to its end, the Chilean government decided
to set up a national observatory and asked Gilliss if they could purchase
the expedition's equipment. Gilliss was very much interested in
a continuation of the work at the observatory in Santiago and he was authorized
by the Secretary of the Navy to sell the instruments, books, etc at the price
paid for them without costs for transportation. As Gilliss wrote in his last
letter from Santiago to Gerling on August 12th 1852
(\cite{GerlingArchive}, Ms.\ 319/288), Moesta was appointed the first director
of the new Chilean national observatory: \textit{''I write but to tell you that
your old pupil Dr.\ Moesta has been appointed Director of the National
Observatory of Chile with a salary of \textdollar{}2000 {\normalfont (pesos)}.
He will have two assistants, and I have no doubt will so conduct the
establishment as to reflect credit on himself and consequently on his
Professor''}. At the same time Moesta became professor of astronomy and
mathematics at the university in Santiago.

\begin{sloppypar}
On August 17th, 1852 Dr.\ Moesta took charge of the Chilean observatory
\cite{Keenan1985}, equipped with the first American
telescope of considerable size and a German meridian circle (\cite{Gilliss1856}
and \cite{Gerling1853}).
The Chilean government continued to promote astronomy; for example, in 1853 the
President sent an expedition to Peru to observe a total eclipse \cite{Gilliss1854};
they also bought a Kessel's clock with a new
self-winding telegraphic register. Moesta held the position till 1865 and then
returned to Germany, where he continued to publish results from his work in
Santiago and served as Consul for Chile.

One interesting incident highlights later developments in Marburg.
In 1882
Friedrich Schwab, mechanic at the Physics Department of the Philipps-Universität
Marburg and self-trained astronomer at the Gerling Observatory Marburg, came to
Punta Arenas in southern Chile as a member of one of the German teams to observe
the Venus transit on December 6th 1882 \cite{Duerbeck2003}. It is possible that Moesta
used his connections to Marburg and to Arthur Auwers, the head of the group,
and his knowledge about Chile to help Schwab acquire this position.
\end{sloppypar}

\subsection{Solar parallax by observation of the opposition of Mars 1862}
\label{subsec_mars_opposition_1862}
In July 1856, when Gilliss and Gould realized the extent of data lost from the
northern observatories, Gilliss started thinking about a new source of 
corresponding measurements. In a letter dated July 7th, 1856
(\cite{GerlingArchive}, letter Ms.\ 319/296) he revealed to Gerling:
\textit{''I have already taken steps to have new differentials made by the
Washington and Santiago {\normalfont (observatories)} at the opposition of
Mars April - June 1858''}! At that time he did not have the power to organize
this new campaign, but by 1859 he was totally occupied with that idea again. He
conveyed to Gerling the content of a letter from Prof.\ Peirce to the President
of the U.S.: \textit{''He {\normalfont (Gilliss)} has demonstrated the
possibility of a new and untried method of its (the solar parallax)
determination, and if the requisite observations had been made by the
astronomers of the northern hemisphere with all the appliances so readily at
hand, with the same fidelity and skill which are manifest in his own
observations, conducted under much difficulty at his temporary station, the
solution would have been all which could be de\-sired.
 \ldots 
How greatly would it \ldots {\normalfont (be for)} the advancement of science
if \ldots (he) were directed to undertake a new determination of the parallax,
with such ample powers and means as would secure the proper cooperation in the
northern hemisphere and enable an American astronomy to have the honor of
permanently establishing the value of the great basis of astronomical
measurements and fixing the vastest of all the recognized standards of
lengths''} (\cite{GerlingArchive}, Ms.\ 319/301, Gilliss to Gerling, May
5th 1859).

However, Gilliss had to wait until in 1861, when he was assigned the head of the
U.S.\ Naval Observatory. In 1862 he announced a campaign to observe the
oncoming Mars opposition \cite{Gilliss1862}, which he organized together with
Moesta in Santiago. He received observations from Santiago, Uppsala, Leiden and
Albany. Reducing those data together with the results from the Washington
observations a new value for the solar parallax could be extracted
\cite{Gilliss1865} \textit{''by the joint activity of  the two observatories 
founded through  his own exertions five thousand  miles apart''} \cite{Gould1866}.

Simon Newcomb, astronomer at the U.S.\ Naval Observatory since 1861,
additionally collected data from observations of the Mars opposition 1862 at
Pulkowa, Helsingfors, Greenwich, Williamstown and Cape of the Good Hope, and
finally adopted a solar parallax of $8\farcs848 \pm 0\farcs013$
\cite{Newcomb1867}, which was a great improvement over the parallax that Encke
had derived from the 18th century observations of the Venus
transits.

\subsection{Contacts and exchange of literature}
\label{subsec_exchange}
A very interesting role in this story was played by Dr.\ Flügel.
As U.S.\ Consul in Germany he acted as a mediator with a
special interest in science! Using his contacts in both the Old and the New
World, he was able to transfer news to the right people in the right place.
Alexander von Humboldt did know that very well and took advantage of this
expedient way to pass information, letters, recommendations etc to the U.S.\
\cite{Schwarz2004}.

\begin{figure}[t]
\begin{center}
\resizebox{0.78\columnwidth}{!}{\includegraphics{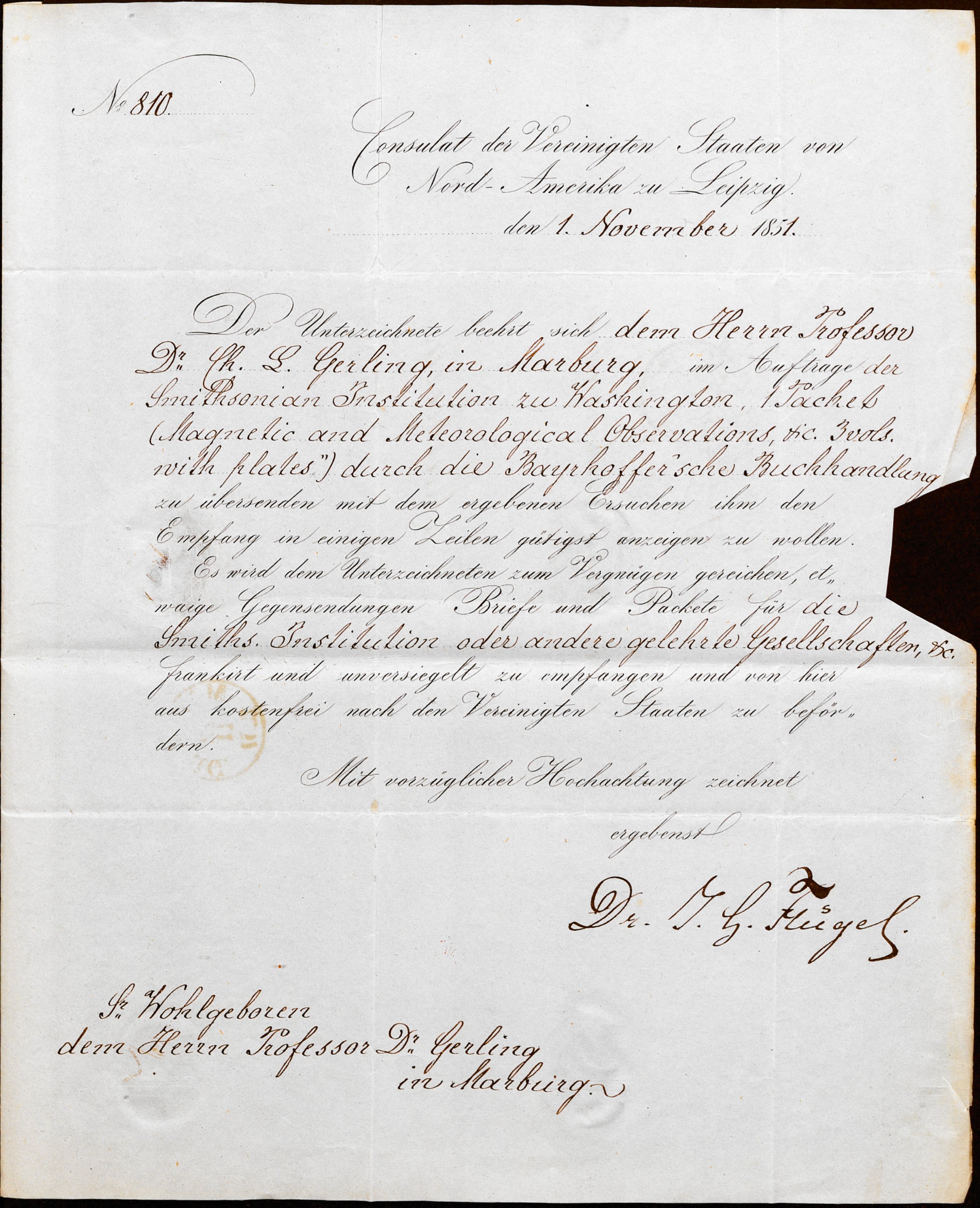}}
\caption{The Smithsonian Institution was pleased to send Gerling a box of
magnetic and meteorological observations and encouraged him, to return in
exchange letters and packages of further materials like essays, books,
scientific reports, etc (\cite{GerlingArchive}, Ms.\ 319/246).}
\label{smithsonian_institution_1}
\end{center}
\end{figure}
 
Dr.\ Flügel probably initiated the contact between Gerling and
Gilliss. During the entire expedition he was circulating information about
its progress. Gerling and Gilliss,  wrote letters
directly to each other, but whenever they wanted the information to be passed
along to the scientific community, they sent it to Flügel! He was the one who
distributed the circular of the expedition, asking observatories to
participate in the project. He also distributed reports from Gilliss and
conveyed any comments he received back to him. 

\begin{figure}[t]
\begin{center}
\resizebox{0.81\columnwidth}{!}{\includegraphics{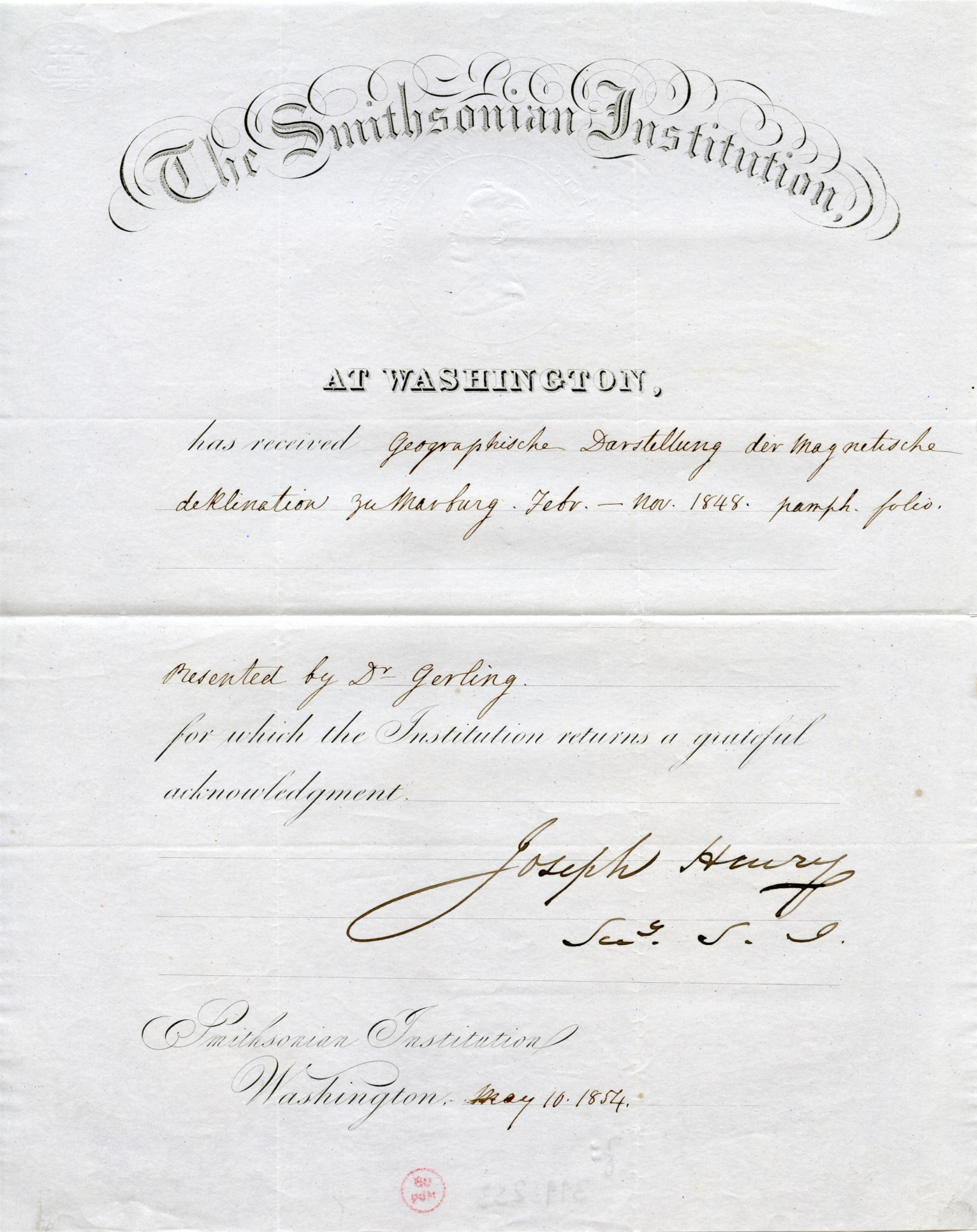}}
\caption{Through Dr.\ Flügel Gerling was asked to send reports to the
Smithsonian Institution. Here Joseph Henry, the ``father'' of the
self-inductance and at that time the secretary of the Smithsonian Institution,
acknowledged the receipt of measurements of the magnetic declination in Marburg 
(\cite{GerlingArchive}, included in Ms.\ 319/253, Flügel to Gerling,
November 8th 1854).}
\label{smithsonian_institution_2}
\end{center}
\end{figure}

Furthermore, in his position as the representative of the Smithsonian
Institution, Flügel organized the transfer of reports, books and other
literature to libraries and scientific institutions
(Figs.\ \ref{smithsonian_institution_1} and \ref{smithsonian_institution_2}).
These were usually sent in boxes and in the cover letters he included the lists
of the material. For example, among the letters Ms.\ 319/222 and Ms.\ 319/223 
from spring 1848 \cite{GerlingArchive} were listed: a perfect facsimile of the
entire Declaration of Independence, documents from the Senate \mbox{Document}
room, a volume of the \textit{Transactions of the American Philosophical Society},
\textit{The Potomac Aqueduct} (Report and Maps), Hassler's comparison of
weights and measures of length and capacity, the \textit{National
Intelligencer} (the most important newspaper in Washington at that time), reports on meteorological and
astronomical observations.

After the results of the expedition had been published, Dr.\ Flügel kept in
touch with Gerling, sending information and seeking advice on various
scientific topics. 
  
\section{Conclusions}
\label{sec_conclusions}
Christian Ludwig Gerling's idea has never really been challenged. The
success of using the Mars oppositions lies in the sufficient reliance in
determining a planet's position by differential measurements with respect to
stars (and not to the sun). The methods surely would have been more successful
if they had been conducted with the same intensity as the Venus transit
measurements in the 19th century. Later, in 1900 and 1931, the improvement of
the star catalogues and the emergence of photography led to a tremendous
increase in accuracy of the solar parallax value by observing the
asteroid Eros (see e.g. \cite{Hinks1904}).
 
James M. Gilliss founded two observatories: the U.S.\ Naval Observatory
after his first experience with continuous astronomical observations and the
National Astronomical Observatory of Chile as a result of the expedition to the
southern hemisphere. 
 
\begin{sloppypar} 
In the 19th century the time was ripe for both scientists’ endeavors. First,
the young U.S.\ community was ready to become involved in challenging scientific
projects. Second, the awareness grew that high quality astronomical
observatories should be equally distributed across the globe, due to that
constantly rotating and moving \mbox{observation} place, we call our earth. 
The foundation of the National Astronomical Observatory of Chile was just the
beginning of a series of new observatories being founded and or supported in
different parts of the world by the Old World \cite{LeGuetTully2013}. 
\end{sloppypar} 
 
Finally, Gerling seems to have made a wise  choice of Valparaiso/Santiago. To
this day, the European Southern Observatory (ESO) is operating some of the most
advanced telescopes of the world in that area.

\begin{acknowledgement}
\textit{Acknowledgements}. I appreciate the cooperation with the library of the
Philipps--Universität Marburg, where the scientific estate of Christian Ludwig
Gerling is stored.
Especially Dr.\ Bernd Reifenberg was very helpful in getting access to the
original letters and materials of Gerling's inheritance.
\end{acknowledgement}

\end{document}